\documentstyle[12pt,epsf]{article}

\begin{document}
\begin{flushright}
IFUP-TH 20/97 \\
IFUM 567/FT
\end{flushright}
\vskip 2cm

\centerline{{\bf Topology at zero and finite $T$ 
in $SU(2)$ Yang-Mills theory}\footnote {Partially 
supported by EC Contract CHEX-CT92-0051 and by MURST.}}
\vskip 5mm
\centerline{B. All\'es\footnote{Address after 1/April/97:
Dipartimento di Fisica, Sezione Teorica, 
Universit\`a degli Studi di Milano and INFN,
Via Celoria 16, 20133-Milano, Italy.}, M. D'Elia and A. Di Giacomo}
\centerline{\it Dipartimento di Fisica dell'Universit\`a and INFN, 
Piazza Torricelli 2, 56126-Pisa, Italy}
\begin{abstract}
We determine the topological susceptibility $\chi$ at $T=0$ and its behaviour 
at finite $T$ across the deconfining transition in pure 
$SU(2)$ gauge theory. We use an improved topological charge density 
operator. $\chi$ goes to zero above $T_c$, but more slowly than
in $SU(3)$ gauge theory.
\end{abstract}

\vfill\eject

\section{Introduction}

The anomalous breaking of the flavour singlet axial symmetry in QCD 
is driven by instanton effects \cite{thoft}. 
This breaking brings about a large mass for the pseudoscalar singlet \cite{witten,veneziano}
\begin{equation}
m^2_{\eta'} = {{2 N_f} \over f^2_\pi} \chi -
m^2_\eta + 2 \; m^2_K .
\label{eq:masseta}
\end{equation}
$\chi$ is the topological susceptibility of the pure gauge theory 
\begin{equation}
 \chi \equiv \int d^4 x \langle 0 | T(Q(x) Q(0))| 0 \rangle_{\rm quenched} ,
\label{eq:chidef}
\end{equation}
with
\begin{equation}
Q(x) = {{g^2} \over {64 \pi^2}} \epsilon^{\mu \nu \rho \sigma} 
 F^a_{\mu \nu} (x) F^a_{\rho \sigma} (x).
\label{eq:qdef}
\end{equation}
The prediction of eq.~(\ref{eq:masseta}) is
$\chi \approx (180 \;\; \hbox{\rm MeV})^4$.

The behaviour of $\chi$ at high temperature $T$ has also physical
relevance (see for instance \cite{shuryak}).
Debye screening in the quark--gluon plasma is expected to produce a 
suppression of the topological susceptibility above the deconfining 
temperature $T_c$ \cite{yaffe}. 

In \cite{delia} we determined, by a numerical simulation on the lattice, $\chi$
at zero and finite temperature in the pure $SU(3)$ gauge theory.
At  $T=0$ we obtained the value $(\chi)^{1/4} = 175(5)$ MeV, which is 
consistent with previous determinations \cite{teper,mplombardo} and with the 
prediction of eq.~(\ref{eq:masseta}). We also showed that $\chi$ 
keeps approximately
constant below $T_c$ and has a
sharp drop at the transition point $T=T_c$. At $T/T_c \approx 1.4$ 
the susceptibility $\chi$ reduces to a few per cent of its value 
before the transition.

In this paper we determine $\chi$ for $SU(2)$ pure gauge theory both at
$T = 0$ and at the deconfining transition with the same technique 
used in ref. \cite{delia}


In  Section~2 we review the method. Our results are presented in Section~3. In section~4 we give some concluding remarks.

\section{The method}

The topological charge was measured on the lattice with the 
operators~\cite{divecchia,haris}
\begin{equation}
Q_L^{(i)}(x) = {{-1} \over {2^9 \pi^2}} 
\sum_{\mu\nu\rho\sigma = \pm 1}^{\pm 4} 
{\tilde{\epsilon}}_{\mu\nu\rho\sigma} \hbox{Tr} \left( 
\Pi^{(i)}_{\mu\nu}(x) \Pi^{(i)}_{\rho\sigma}(x) \right).
\label{eq:qlattice}
\end{equation}
${\tilde{\epsilon}}_{\mu\nu\rho\sigma}$ is the
standard Levi-Civita tensor for positive directions while for negative
ones the relation ${\tilde{\epsilon}}_{\mu\nu\rho\sigma} =
- {\tilde{\epsilon}}_{-\mu\nu\rho\sigma}$ holds. 
$\Pi^{(i)}_{\mu\nu}$ is the plaquette in the $\mu - \nu$ plane
constructed with $i$-times smeared links $U_\mu^{(i)}(x)$.
These smeared links are defined recursively starting from the
ordinary link $U_\mu(x)$ as 
\begin{eqnarray}
U^{(0)}_{\mu}(x) &=& U_{\mu}(x), \nonumber \\
{\overline U}^{(i)}_{\mu}(x) &=& (1-c) U^{(i-1)}_{\mu}(x) +
{c \over 6} 
\sum_{{\scriptstyle \alpha = \pm 1} \atop { \scriptstyle 
|\alpha| \not= \mu}}^{\pm 4}
U^{(i-1)}_{\alpha}(x) U^{(i-1)}_{\mu}(x+\hat{\alpha})
U^{(i-1)}_{\alpha}(x+\hat{\mu})^{\dag}, \nonumber \\
U^{(i)}_{\mu}(x) &=& {{{\overline U}^{(i)}_{\mu}(x)} \over
{ \left( {1 \over 2} \hbox{Tr} {\overline U}^{(i)}_{\mu}(x)^{\dag} 
{\overline U}^{(i)}_{\mu}(x) \right)^{1/2} } }.
\label{eq:qsmeared}
\end{eqnarray}
The parameter $c$ can be tuned in order to optimize the improvement of 
the operator. We choose $c=0.85$ and measure the topological charge
for $i=$0, 1, 2.

The topological susceptibility from the $i$-smeared operator 
on the lattice is calculated by
\begin{equation}
\chi^{(i)}_L = \langle \sum_x Q_L^{(i)} (x) 
Q_L^{(i)} (0) \rangle.
\label{eq:chilattice}
\end{equation}
$\chi^{(i)}_L$ mixes with the continuum susceptibility $\chi$ and with all 
renormalization group invariant operators of dimension $\leq 4$, i.e.
the trace of the energy-momentum tensor and 
the unity operator \cite{campo329}. The mixing to continuum $\chi$ is 
described by the square of the (finite) multiplicative renormalization of the 
topological charge operator $Q_L^{(i)}$ to the continuum operator $Q$ 
\cite{z1}:
\begin{equation}
Q_L^{(i)} = Z^{(i)}(\beta ) Q a^4 + {\cal O}( a^6 ).
\end{equation}
Therefore the following relation holds
\begin{equation}
\chi_L^{(i)}  = Z^{(i)}(\beta)^2 a(\beta)^4 \chi + M^{(i)}(\beta)
+ O(a^6),
\label{eq:chimixings}
\end{equation}
where $M^{(i)}(\beta)$ is the mixing to the trace of the energy-momentum 
tensor and to the unity operator. 
As usual $\beta\equiv 2N_c/g^2$ where $N_c$ is the number of colours
and $g$ the gauge coupling.

The additive renormalization $M^{(i)}(\beta)$ can be 
determined by thermalizing the short range fluctuations starting from 
a zero field configuration, without changing the (zero) topological
content of it ~\cite{vicari,gunduc,npbproc,farchioni,delia}. 
We start with the flat configuration
(all links $U_\mu(x)=1$) and create a sample of configurations
by applying some heat--bath steps. At each step we measure the topological
susceptibility.
The content of instantons is checked on intermediate steps, by cooling a copy 
of the configuration. Configurations where instantons or antiinstantons 
have been produced are eliminated from the sample.
A plateau is reached after   
$\sim 10-20$ steps that keeps constant along $\sim 100$ updating
steps. The value for $M^{(i)}(\beta)$
is the average of this plateau on the ensemble.
The number of discarded
configurations depends on the value of $\beta$. At $\beta = 2.5$ the rate 
of discarded configurations per heating step is $\sim1$\%; at $\beta = 2.6$
it drops to $\sim 0.5$\% and is well below $0.1$\% at $\beta = 2.7$.

By subtracting $M^{(i)}(\beta)$ we naturally 
impose the condition that the physical topological susceptibility 
be zero on topologically trivial configurations, thus matching
the continuum renormalization prescription for $\chi$.

The multiplicative renormalization $Z^{(i)}(\beta)$ 
can be determined in a similar way.
We thermalize short range fluctuations starting from a 1 instanton 
configuration.
At each updating step we measure the
topological charge. Again checks are performed to eliminate
configurations where the topological content of the starting 
configuration has been changed 
during the updating procedure. 
$Q_L^{(i)}$ is then measured: it stabilizes on a plateau where 
$Q_L^{(i)} = Z^{(i)}(\beta ) Q$, and the value of $Z^{(i)}(\beta )$ 
is extracted from 
the average on these plateaux.

The statistical accuracy in the determination of the physical value of $\chi$
strongly depends on  $M^{(i)}(\beta)$ and $Z^{(i)}(\beta)$, which in their
turn depend on the lattice regularization $Q_L^{(i)}$ used for the topological 
charge. The improvement of the operator results in a better accuracy 
\cite{haris,delia}.


\section{Results}

 The zero temperature determination of $\chi$ has been done on
a $16^4$ lattice at three different values of $\beta$ with the Wilson action.
Statistical errors have been estimated by
using a standard blocking 
procedure. 
To fix the scale of length we refer to ref.  \cite{karsch}. We put
\begin{equation}
a(\beta) = {1 \over \Lambda_L} \lambda(\beta) f(\beta)
\label{eq:scaling_k}
\end{equation}
where $f(\beta)$ is the usual 2--loop scaling function and $\lambda(\beta)$
is a corrective factor tabulated in Table 1 of 
ref. \cite{karsch}, which tends to one at large $\beta$, where asymptotic 
scaling sets up.  

 The results for $\chi/\Lambda_L^4$ at zero temperature for 
0,1,2--smearings are shown in Table~I and Figure~1. 
Scaling is observed in each of them.
The three determinations (0,1,2-smearings) agree
within errors. The smearing process has indeed improved
the result by drastically reducing the error bars. 

To convert to physical units we need a determination of 
$\Lambda_L$. We get it by using the result of ref. \cite{karsch} 
${T_c/\Lambda_L} = 21.45(14)$, and that of ref. \cite{fingberg} 
${T_c/\sqrt{\sigma}} = 0.62(2)$ ($\sigma$ is the string tension). 
The result is $\Lambda_L = 14.15(42)$ MeV, where, as usual, we have assumed
$\sqrt{\sigma} = 440$ MeV. 
Combining the \break 2--smeared results at the three $\beta$ values 
we obtain
$(\chi)^{1/4} = (198 \pm 2 \pm 6)$ MeV, 
where the first error comes from our determination
and the second from the uncertainty on $\Lambda_L$. 

We have made a comparison with existing determinations of $\chi$
at~$T=0$. We agree with ref. \cite{deforcrand} where
$(\chi)^{1/4}$=200(15) MeV is obtained by use of improved cooling.
In ref. \cite{boulder} 230(30) MeV is 
quoted for the same quantity, which is computed
by a different (improved) action and with an improved geometric algorithm.
The result of refs. \cite{campo329,giannetti}, 
when converted in MeV by use of the same scale as in this paper, is
$(178\pm 1\pm 5)$~MeV. However there the renormalization constants were computed
by perturbation theory because the method of ref. \cite{vicari} did not
exist yet. In ref. \cite{teper2} $\chi\approx 130$~MeV 
is obtained which is lower; we are not able to trace back what part of the 
difference is due to the different method of computing $(\chi)^{1/4}$
(cooling) and what part comes from the determination of the scale $a$.
As for ref. \cite{schierholz} the comparison will be done systematically
in a forthcoming paper \cite{kirchner}.

At finite temperature we used a $32^3\times 8$ lattice: 
at this size 
the deconfining transition 
is located at $\beta_c = 2.5115(40)$~\cite{fingberg}.
The temperature $T$ as a function of $\beta$ is given by
\begin{equation}
T = {1 \over {N_{\tau} a(\beta)}}. 
\label{eq:temperature}
\end{equation}
The results for 1,2--smearings are shown in Table~II and figure~2: the data 
for the 0--smeared operator have very large errors above the 
deconfining temperature 
and are not shown in the figure. 

At $T < T_c$ our data are consistent with the value
at zero temperature, while a 
drop is observed above the deconfining transition. This behaviour
was also observed in the $SU(3)$ gauge theory \cite{delia}; however
in the $SU(3)$ case the drop is quite steeper than in the present 
case. This qualitative difference between the two gauge theories 
can be well appreciated in Figure~3, where $\chi/\chi_{(T = 0)}$ 
is plotted versus $T/T_c$ for both 
$SU(2)$ and $SU(3)$.

\section{Concluding remarks}

We have determined 
the topological susceptibility of $SU(2)$ pure gauge theory 
at zero temperature and its behaviour through $T_c$. 
The improvement of the topological charge density operator \cite{haris} 
has made this determination possible.

If, to fix the scale, the $SU(2)$ string tension is assumed to be the physical
one, $\chi_{(T = 0)}$ results slightly larger than for $SU(3)$. $\chi$ is 
approximately 
constant below $T_c$, and drops to zero above the transition, however 
more slowly than for $SU(3)$.

\section{Acknowledgements}

We are greatly indebted to CNUCE (Pisa) for qualified technical 
assistance in the use of their IBM--SP2 and to the CRT Computer Center of 
ENEL (Pisa) for warm hospitality and collaboration in the use of their 
Cray YMP--2E.

\vskip 15mm


\newpage

\noindent{\bf Figure captions}

\begin{enumerate}

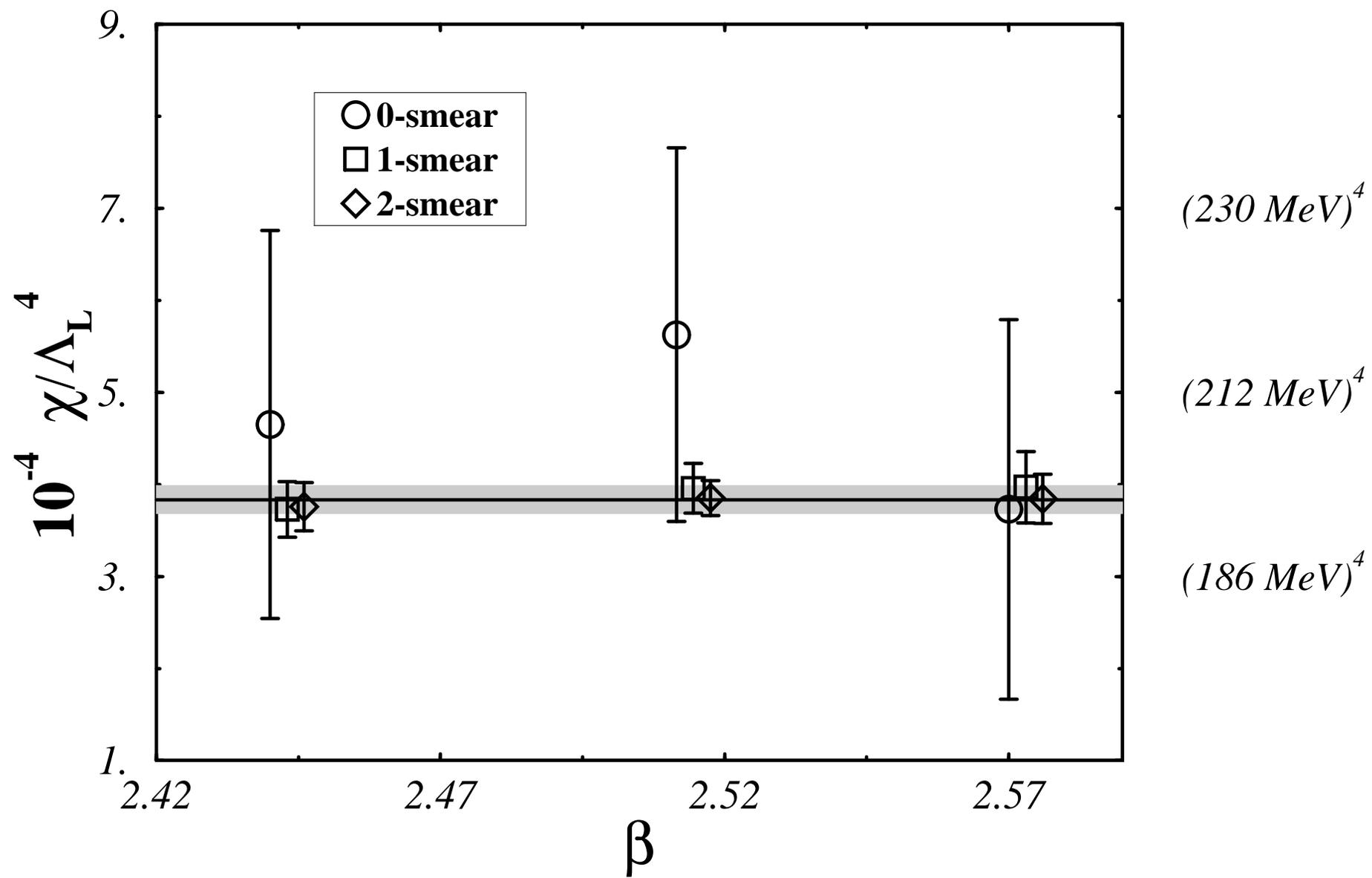
\item[Figure 1.] $\chi$ at $T=0$.
The straight line is the result of the linear fit of the 2-smeared data.
The improvement from $Q^{(0)}_L$ to $Q^{(2)}_L$ is clearly visible.
$\Lambda_L$=14.15 MeV was used to fix the scale by eq. (9).

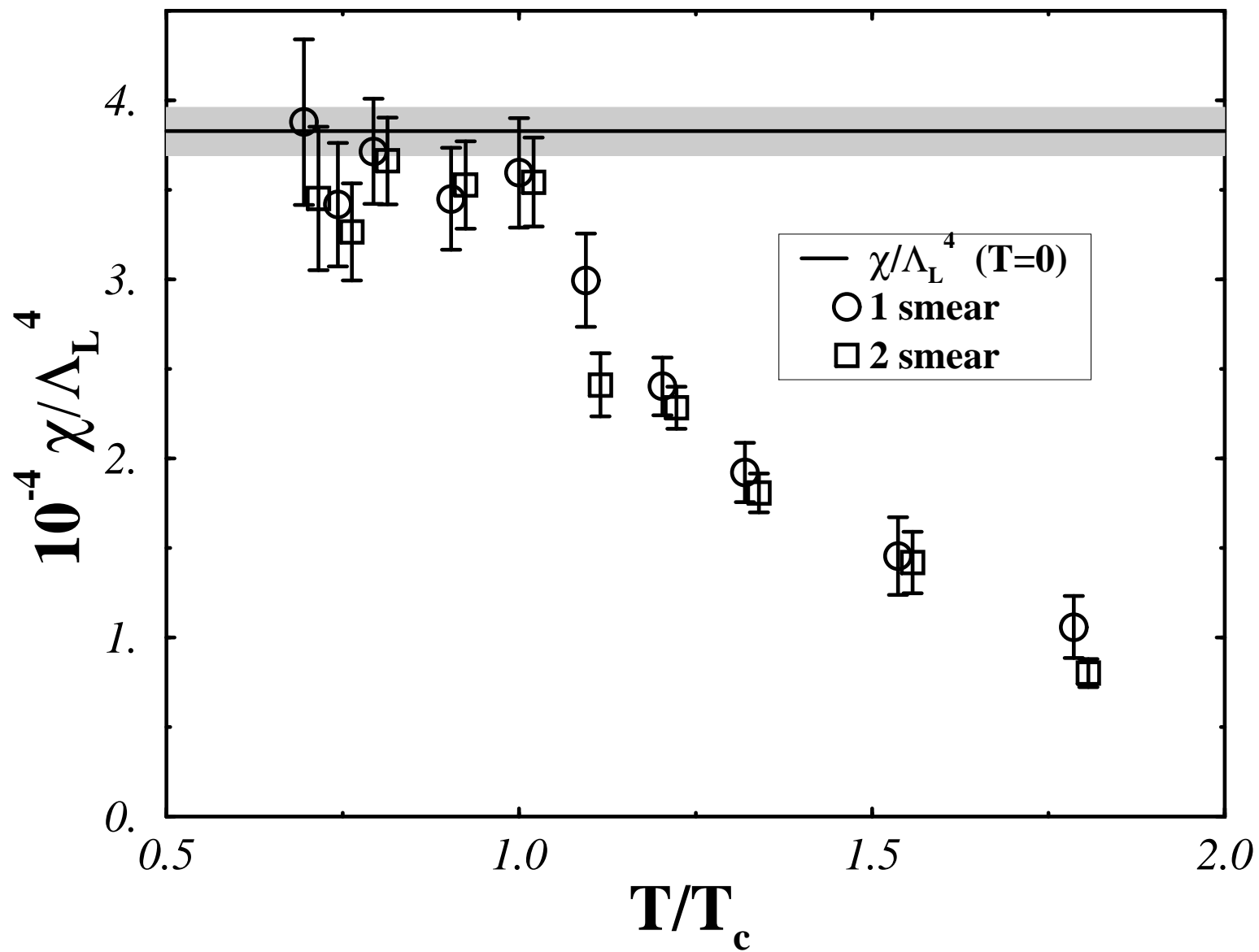
\item[Figure 2.] $\chi/\Lambda_L^4$ versus $T/T_c$ across
the deconfining phase transition for 1 and 2 smearings.
The horizontal band is the determination at $T=0$ of Figure~1.

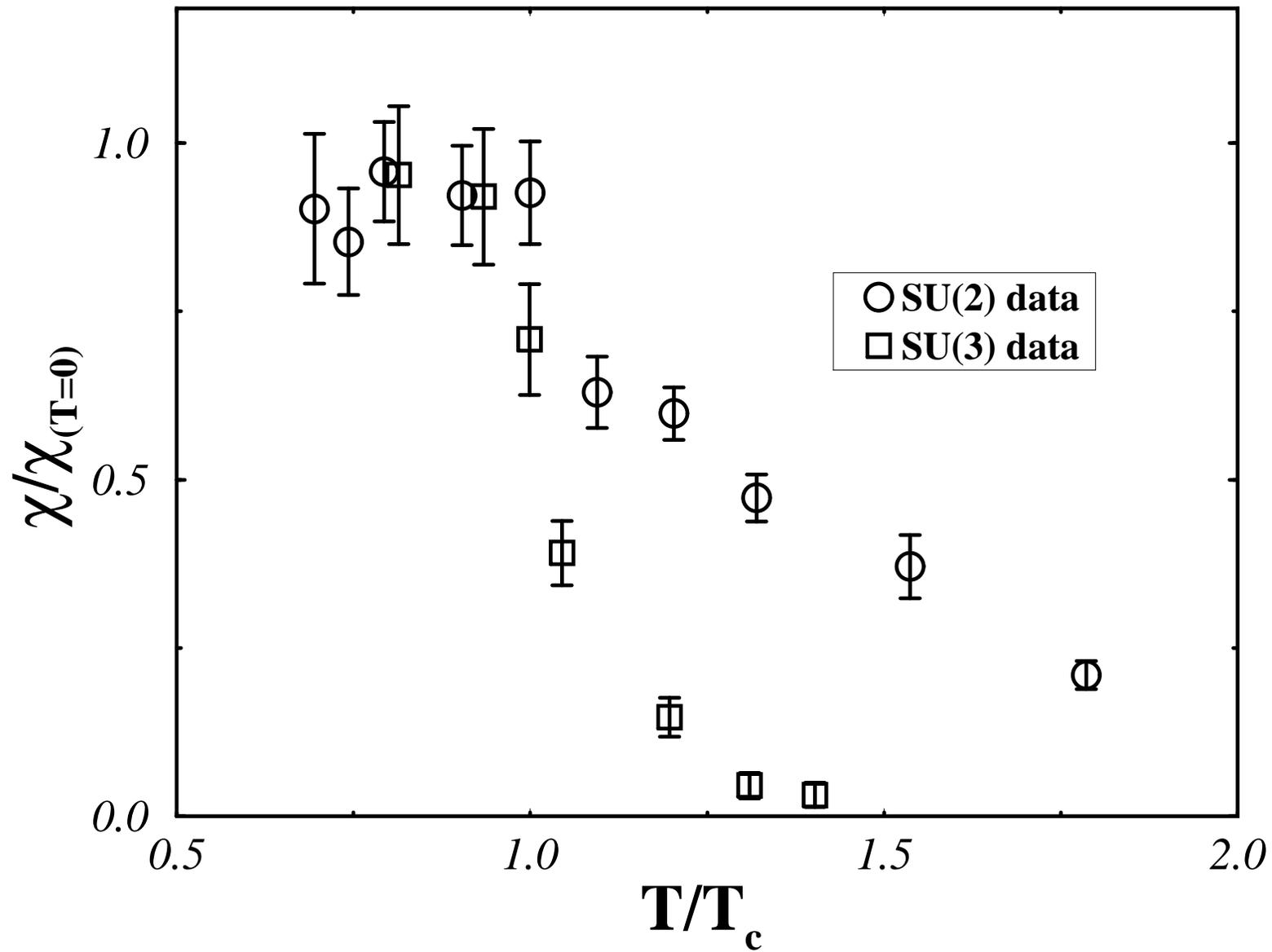
\item[Figure 3.] The ratio $\chi/\chi_{(T=0)}$ as a function of 
$T/T_c$ for $SU(2)$ and $SU(3)$ for the 2-smeared data.

\end{enumerate}

\vskip 2cm

\noindent{\bf Table captions}

\vskip 5mm

\begin{enumerate}

\item[Table I.] $\chi/\Lambda_L^4$ from the 0,1 and 2-smeared operators
at $T=0$. $\chi^{(i)}$ means the continuum susceptibility obtained
from the $i$-smeared operator.

\item[Table II.] $T/T_c$, $\chi^{(1)}/\Lambda_L^4$, $\chi^{(2)}/\Lambda_L^4$
as a function of $\beta$. Same notation as in Table~I.

\end{enumerate}

\newpage

\vskip 2cm

\centerline{\bf Table I}

\vskip 5mm

\moveright 0.1 in
\vbox{\offinterlineskip
\halign{\strut
\vrule \hfil\quad $#$ \hfil \quad & 
\vrule \hfil\quad $#$ \hfil \quad & 
\vrule \hfil\quad $#$ \hfil \quad & 
\vrule \hfil\quad $#$ \hfil \quad \vrule \cr
\noalign{\hrule}
\beta & 
10^{-4} \times \chi^{(0)}/\Lambda_L^4 &
10^{-4} \times \chi^{(1)}/\Lambda_L^4 &
10^{-4} \times \chi^{(2)}/\Lambda_L^4 \cr
\noalign{\hrule}
2.44 & 4.7(2.1) & 3.73(30) & 3.76(26) \cr
\noalign{\hrule}
2.5115 & 5.6(2.0) & 3.96(27) & 3.85(19) \cr
\noalign{\hrule}
2.57 & 3.7(2.1) & 3.97(39) & 3.84(27) \cr
\noalign{\hrule}
}}

\vskip 2cm

\centerline{\bf Table II}

\vskip 5mm

\moveright 0.4 in
\vbox{\offinterlineskip
\halign{\strut
\vrule \hfil\quad $#$ \hfil \quad & 
\vrule \hfil\quad $#$ \hfil \quad & 
\vrule \hfil\quad $#$ \hfil \quad & 
\vrule \hfil\quad $#$ \hfil \quad \vrule \cr
\noalign{\hrule}
\beta & 
T/T_c &
10^{-4} \times \chi^{(1)}/\Lambda_L^4 &
10^{-4} \times \chi^{(2)}/\Lambda_L^4 \cr
\noalign{\hrule}
2.40 & 0.695 & 3.88(46) & 3.45(40) \cr
\noalign{\hrule}
2.42 & 0.743 & 3.42(34) & 3.26(27) \cr
\noalign{\hrule}
2.44 & 0.793 & 3.71(29) & 3.66(24) \cr
\noalign{\hrule}
2.48 & 0.904 & 3.45(28) & 3.53(24) \cr
\noalign{\hrule}
2.5115 & 1.000 & 3.60(31) & 3.54(25) \cr
\noalign{\hrule}
2.54 & 1.095 & 2.99(26) & 2.41(18) \cr
\noalign{\hrule}
2.57 & 1.203 & 2.40(16) & 2.28(12) \cr
\noalign{\hrule}
2.60 & 1.320 & 1.92(17) & 1.81(11) \cr
\noalign{\hrule}
2.65 & 1.538 & 1.46(22) & 1.42(17) \cr
\noalign{\hrule}
2.70 & 1.786 & 1.06(17) & 0.83(8) \cr
\noalign{\hrule}
}}

\end{document}